\documentclass[slac_one]{revtex4}
\usepackage{graphicx}
\usepackage{fancyhdr}
\newcommand{\eqref}[1]{Eq.~(\ref{#1})}

\begin{document}

\title{{\small{LEPTON - PHOTON 2007}}\\ 
\vspace{12pt}
Indirect signals from braneworlds} 

\author{Jose~A.~R.~Cembranos}
\affiliation{Department of Physics and Astronomy, University of
California, Irvine, CA 92697, USA}

\author{\'Alvaro~de~la~Cruz-Dombriz}
\affiliation{Departamento de F\'{\i}sica Te\'orica I,
Universidad Complutense de Madrid, 28040 Madrid, Spain}

\author{Antonio~Dobado}
\affiliation{Departamento de F\'{\i}sica Te\'orica I,
Universidad Complutense de Madrid, 28040 Madrid, Spain}

\author{Antonio~L.~Maroto}
\affiliation{Departamento de F\'{\i}sica Te\'orica I,
Universidad Complutense de Madrid, 28040 Madrid, Spain}

\begin{abstract}
It has been suggested that our universe could be a
3-dimensional brane where the SM fields live embedded in a
D-dimensional space-time.  In flexible braneworlds, 
in addition to the SM fields, new degrees of freedom appear on the brane
associated to brane fluctuations. These new fields, known as branons, are
standard WIMPs (Weakly Interacting Massive Particles) and therefore natural dark matter candidates, whose spontaneous
annihilations can provide first evidences for this scenario.
\end{abstract}

\maketitle


\thispagestyle{fancy}

\section*{Introduction}
\label{aba:sec1}

Dark matter (DM) is one of the most intriguing puzzles in physics. 
The fact that DM cannot be made of any of the known 
particles is one of the most appealing arguments for the existence of new physics.
The experimental search for its nature needs 
the interplay of new collider experiments \cite{colliders} and 
astrophysical observations. These last ones use to be classified 
in direct or indirect searches (read \cite{Feng:2003xh,structureformation} 
however, for different alternatives). Elastic scattering of DM particles
from nuclei should lead directly to observable nuclear recoil signatures.
On the other hand, dark matter might be detected indirectly, by observing 
the products of their annihilation into standard model (SM) particles. 
We will focus our discussion on this last alternative.

\begin{table}
\centering \small{
\begin{tabular}{||c|cccc||}
\hline Experiment
&
$\sqrt{s}$(TeV)& ${\mathcal
L}$(pb$^{-1}$)&$f_0$(GeV)&$M_0$(GeV)\\
\hline
%
%
HERA$^{\,1}
$& 0.3 & 110 &  16 & 152
\\
Tevatron-I$^{\,1}
$& 1.8 & 78 &   157 & 822
\\
Tevatron-I$^{\,2}
$ & 1.8 & 87 &  148 & 872
\\
LEP-II$^{\,2}
$& 0.2 & 600 &  180 & 103
\\
\hline
Tevatron-II$^{\,1}
$& 2.0 & $10^3$ &  256 & 902
\\
Tevatron-II$^{\,2}
$& 2.0 & $10^3$ &   240 & 952
\\
ILC$^{\,2}
$& 0.5 & $2\times 10^5$ &  400 & 250
\\
ILC$^{\,2}
$& 1.0 & $10^6$ &  760 & 500
\\
LHC$^{\,1}
$& 14 & $10^5$ &  1075 & 6481
\\
LHC$^{\,2}
$& 14 & $10^5$ &   797 & 6781
\\
CLIC$^{\,2}
$& 5 & $10^6$ &  2640 & 2500
\\
\hline
\end{tabular}
} \caption{\footnotesize{Limits from 
direct branon searches in colliders (results at the $95\;\%$ c.l.). 
Upper indices $^{1,2}$ denote 
monojet and single photon channels respectively. 
The results for 
HERA, LEP-II and Tevatron run I have been obtained
from real data, whereas those for Tevatron run II, 
ILC, LHC and CLIC  are estimations.
$\sqrt{s}$ is the center of mass energy of the total
process; ${\mathcal L}$ is the total integrated luminosity;
 $f_0$ is the bound on the brane tension scale for one
massless branon ($N=1$) and $M_0$ is the limit on the branon mass for 
small tension $f\rightarrow0$ (see \cite{ACDM} for details).}}
\label{tabHad}
\end{table}

It has been found that massive brane fluctuations (branons)
are natural candidates to dark matter in brane-world models
with low tension \cite{CDM}. From the point of view of the 4-dimensional 
effective phenomenology, the massive branons are new 
pseudoscalar fields which can be understood as the
pseudo-Goldstone  bosons corresponding to the spontaneous 
breaking of translational
 invariance in the bulk space produced by the presence of the 
brane \cite{BR, ACDM}. They are
 stable due to  parity invariance on the brane. 
The  SM-branon low-energy effective Lagrangian 
\cite{BR, ACDM}  can be written as

\begin{eqnarray}
{\mathcal L}_{Br}=
\frac{1}{2}g^{\mu\nu}\partial_{\mu}\pi^\alpha
\partial_{\nu}\pi^\alpha-\frac{1}{2}M^2\pi^\alpha\pi^\alpha
+
\frac{1}{8f^4}(4\partial_{\mu}\pi^\alpha
\partial_{\nu}\pi^\alpha-M^2\pi^\alpha\pi^\alpha g_{\mu\nu})
T^{\mu\nu} ,
\,\label{lag}
\end{eqnarray}
where $\alpha=1\dots N$, with $N$ the number of branon
species. 

We see that branons interact by pairs with the SM
energy-momentum tensor $T^{\mu\nu}$, and that the coupling
is suppressed by the brane tension $f^4$. 
Limits on the model parameter from tree-level processes in colliders  
are briefly summarized in Table \ref{tabHad}, where not 
only present restrictions coming from HERA, Tevatron and LEP-II are provided,  
but also some
prospects for future colliders such as ILC, LHC or CLIC \cite{ACDM,L3,CrSt}.
Additional bounds from astrophysics and 
cosmology can be found in \cite{CDM}.

\begin{figure}[bt]
\begin{center}
\resizebox{8.8cm}{6.4cm} 
{\includegraphics{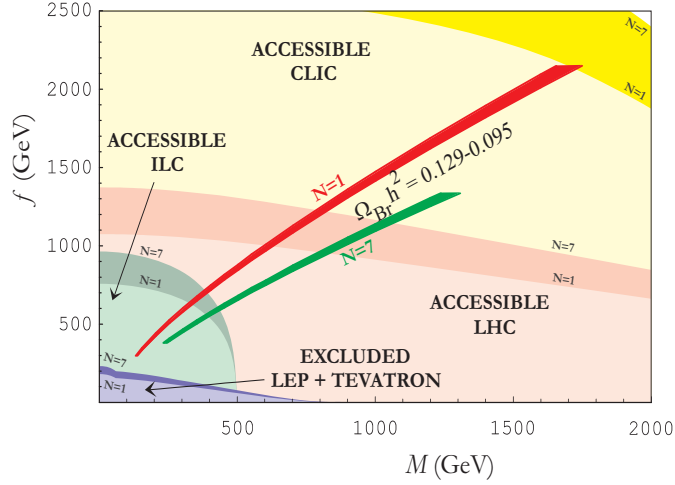}}
\caption {\footnotesize Branon abundance in the range:
$\Omega_{Br}h^2=0.129 - 0.095$, in the $f-M$ 
plane (see \cite{CDM} for details). The regions are only plotted
for the preferred values of the brane tension scale $f$ for the muon anomalous 
magnetic moment \cite{Rad}. 
The lower area is excluded by single-photon processes at LEP-II
together with monojet signals at Tevatron-I
\cite{ACDM,L3}. Prospects for the sensitivity in collider 
searches of real branon production 
are also plotted for future experiments (See \cite{ACDM} and Table \ref{tabHad}).
In this figure one can observe explicitly the dependence on the branons number $N$,
since all these regions are plotted for the extreme values $N=1$ and $N=7$.
} 
\label{CDM}
\end{center}
\end{figure}

Even if branons are stable, two of them may annihilate into ordinary matter such as quarks, leptons and gauge bosons. Their annihilation in different places (galactic halo, Sun, Earth, etc.) produce cosmic rays to be discriminated through distinctive signatures from the background.

After branon annihilation in mentioned particles, a cascade process would occur and in the end, particles to be observed would be neutrinos, gamma rays, positrons and antimatter (antiprotons, antihelium, antideuterions, etc.) that may be detectable through different devices. From those, neutrinos and gamma rays have the advantage of maintaining their original direction. On the contrary, charged antimatter searches are hindered by propagation trajectories.

\section*{Photons}

The photon flux in the direction of the galactic center coming
from dark matter annihilations can be written as \cite{BUB,FMW}:
\begin{eqnarray}
\frac{\text{d}\,\Phi_{\gamma}^{DM}}{\text{d}\,\Omega\,\text{d}\,E_{\gamma}} =
\frac{J_0}{N M^2}\sum_i\langle\sigma_i v\rangle
\frac{\text{d}\,N_\gamma^i}{\text{d}\,E_{\gamma}} ,
\label{flux}
\end{eqnarray}
where $J_0$ is the integral of the dark matter mass density
profile, $\rho(r)$, along the path between the galactic center and
the gamma ray detector:
\begin{eqnarray}
J_0= \frac{1}{4\pi}\int_{path} \rho^2\,\text{d}\,l\, ,
\end{eqnarray}
with $\langle\sigma_i v\rangle$ the thermal average of the
annihilation cross section of two dark matter particles into any other
two particles. On the other hand,  the continuum photon
spectrum from the subsequent decay of particles species $i$ presents a
simple description in terms of the photon energy normalized to the
dark matter mass, $x=E_{\gamma}/M$. Thus, for each channel $i$, we have
\begin{eqnarray}
\frac{\text{d}\,N_\gamma^i}{\text{d}\,x}= M\frac{\text{d}\,N_\gamma^i}{\text{d}\,E_{\gamma}}=
\frac{a^i}{x^{3/2}}e^{-b^ix}\, ,
\end{eqnarray}
where $a^{i}$ and $b^{i}$ are constants. 
In the case of heavy branons, neglecting three
body annihilations and direct production of two photons, the 
main contribution to the photon flux comes from branon annihilation into 
$ZZ$ and $W^+ W^-$. The contribution from heavy fermions, i.e. annihilation
in top-antitop can be shown to be subdominant \cite{indirect}.
The concrete
values for the above constants in those channels are: 
$a^{ZZ}=a^{W^\pm W^\mp}=0.73$
and $b^{ZZ}=b^{W^\pm W^\mp}=7.8$ \cite{BUB,FMW}. 

On the other hand, the thermal averaged cross-section $\langle\sigma_{Z,W} v\rangle$
which enters in eq. (\ref{flux}) has been
calculated in \cite{CDM} and in the non-relativistic limit is given by
\begin{eqnarray}
\langle\sigma_{Z,W} v\rangle=\frac{M^2\sqrt{1-\frac{m_{Z,W}^2}{M^2}}
(4M^4-4M^2m_{Z,W}^2+3m_{Z,W}^4)}{64f^8\pi^2} .
\end{eqnarray}

These annihilations would  produce a broad energy distribution of photons 
centered around $M/10$, which would be difficult to be distinguished from background. 
However, the directional dependence for the gamma ray intensity coming from these 
annihilations can provide a distinctive signature. 

The produced high-energy gamma photons could be in the range
(30 GeV-10 TeV), detectable 
by Atmospheric
Cerenkov Telescopes (ACTs)  such as $\text{HESS}$ \cite{HESS}, $\text{VERITAS}$ or $\text{MAGIC}$. 
On the contrary, if $M<m_{Z,W}$, the
annihilation into W or Z bosons is kinematically forbidden and it is
necessary to take into account the rest of  channels, mainly 
annihilation into the heaviest possible quarks \cite{Silk}. 
In this case, the photon fluxes would be in the range detectable by 
space-based gamma ray observatories \cite{indirect} 
such as $\text{EGRET}$ \cite{EGRET}, $\text{GLAST}$ or $\text{AMS}$, 
with better sensitivities around 30 MeV-300 GeV. 

On the other hand the cross section associated to the three body annihilation 
$\sigma_{\gamma} : \pi(p_1)\pi(p_2) \longrightarrow e^{+}(k_1)e^{+}(k_2)\gamma_{\mu}(q)$ could be important for high energetic photons. The form of this differential cross section is

\begin{eqnarray}
\langle\frac{\text{d}\sigma_{\gamma}}{\text{d}x}v\rangle\,=\,\frac{\alpha M^{6} x [1-\frac{m_{e}^2}{M^{2}(1-x)}]^{1/2}}{90 f^{8} \pi^{2} N}\Big\{x^{2}[9(1-x)+4x^{2}]+3(1-x)[1+(1-x)^{2}]\text{ln}\frac{M^{2}(1-x)}{m_{e}^2}\Big\}\, ,
\end{eqnarray}
where again $x\,=\,E_{\gamma}/M$.

Finally we point out that there would be a gamma ray line from direct annihilation into photons since branons couple directly to them, producing a monochromatic signal at an energy equal to the branon mass. This annihilation is suppressed by $d$-wave but it is a background free signal.

\section*{Neutrinos}

Searches for high energy neutrinos could be done pointing to the Sun, Earth or galactic center. These sources are favoured since branon density is likely enhanced by gravitational capture. For example, neutrinos coming from cascades of branon annihilation inside the Earth, interact in high energy neutrinos telescopes, producing up-going muons fluxes that may be distinguished from down-going ones coming from the atmospheric cosmic ray interactions, which constitute the main physical background. 

The total number of muons per unit area and time above an energy threshold $E_{thr}$ and within a cone of half angle $\theta_{c}$ for the $\text{AMANDA}$ telescope can be evaluated as follows \cite{Amanda}: 

\begin{eqnarray}
\phi^{\text{DM}}_{\mu} (E_{\mu}\geq E_{thr},\theta \geq \theta_{c}) \,=\, \frac{\Gamma_{A}}{4\pi R_{\oplus}^2}\int_{E_{thr}}^{\infty}\text{d}E_{\mu}\int_{\theta_c}^{\pi}\text{d}\theta\frac{\text{d}^2N_{\mu}}{\text{d}E_{\mu}\text{d}\theta}\,,
\label{muon_flux}
\end{eqnarray}
where $R_{\oplus}$ is the Earth radius and the term $\text{d}^2N_{\mu}/\text{d}E_{\mu}\text{d}\theta$ represents the number of muons per unit angle and energy produced from branon annihilation. $\Gamma_{A}$ is the annihilation rate of branons in the center of the Earth and is related to neutrino-to-muon conversion rate $\Gamma_{\nu\mu}$ through 

\begin{eqnarray}
\Gamma_{\nu\mu}(M)\,=\,\Gamma_{A}\frac{1}{4\pi R_{\oplus}^2}\int_{0}^{M}\sum_{i} B_{\text{i}}\big(\frac{\text{d}N^{\text{i}}_{\nu}}{\text{d}E_{\nu}}\big)
\times \sigma_{\nu+N\longrightarrow \mu+...}(E_{,\nu};E_{\mu}\geq E_{thr})\rho_{N}\text{d}E_{\nu}\,.
\label{muon_neutrino_rate}
\end{eqnarray}

The term inside the integral takes into account the muons production through neutrino-nucleon cross sections $\sigma_{\nu+N}$ weighted by the different branching ratios and the corresponding neutrino energy spectra $B_{\text{i}}
\text{d}N^{\text{i}}_{\nu}/\text{d}E_{\nu}$. $\rho_{N}$ is the nucleon density of ice. An upper limit can be given to $\Gamma_{\mu\nu}$, ie. to the number of muons with an energy above the detector threshold produced by neutrino interactions per unit volume and time:

\begin{eqnarray}
\Gamma_{\mu\nu}\leq \frac{N_{\beta}}{V_{eff}t}\,,
\end{eqnarray}
with $V_{eff}$, the effective volume for the detector, and $N_{\beta}$, an upper limit on the signal at a chosen confidence level $\beta\%$  obtained from the number of observed events and the expected atmospheric neutrino background.

As it happens for photons, the main production source of neutrinos is through gauge boson annihilation if such channel is available, or heavy quarks in the other case. The former will produce a broad energy distribution centered around $M/2$. The possibility of mono-energetic neutrinos direct production  with an energy of $M$ is again either suppresed by $d$-wave or by the neutrino mass (for the $s$ and $p$-wave annihilation).

\section*{Positrons}

Despite the uncertainties associated with the background spectrum for positrons, experiments performed with balloons, such as $\text{HEAT}$ collaboration \cite{HEAT_collaboration}, seem to have observed an excess on the positrons continuum. Nevertheless the background flux of positrons, expressed as a fraction of the flux of electrons, decreases slowly with energy, which means that at low energies positrons coming from branon annihilation are expected to be enhanced.

The positron flux generated by branons may be expressed as follows
\begin{eqnarray}
\frac{\text{d}\Phi_{e^{+}}^{\text{DM}} }{\text{d}\Omega  \text{d}\text{E}_{e^{+}}}\,=\,\frac{\rho_{0}^2}{N M^2}\sum_{i}\langle \sigma_{i}v\rangle B_{e^{+}}^{i}\int \text{d}E f_{i}(E)G(E,E_{+})\, ,
\label{diff_cross_section_positrons}
\end{eqnarray}
where $\rho_{0}$ is the local branon mass density, and $N$ is the number of branons with mass $M$. $\langle\sigma_{i}v\rangle$ is the thermal average of the annihilation cross section of two branons into two particles ($i^{th}$ channel) times the relative velocity, $B_{e^{+}}^{i}$ are the branching fractions to positron in channel $i^{th}$. For instance channels $ZZ$ and $W^{\pm}W^{\mp}$ have branching fractions $B_{e^{+}}^{ZZ}\,=\,0.067$ and $B_{e^{+}}^{W^{\pm}W^{\mp}}\,=\,0.11$.

Distortion in the flux due to positron propagation in the galaxy will be accounted by the above integral: $f_{i}(E)$ will describe the initial positron energy distribution from branon annihilation and the Green function, $G(E,E_{e^{+}})$, accounts for distortion itself due to positron propagation. 

The total behavior of the signal may be parametrized as a two-parameters function: $\{M, f(N/{\rho^2})^{1/8}\}$. For branons heavier than gauge bosons, their annihilation into these particles is the dominant source of positrons through gauge boson two-body decays, which produce a positron spectrum peaked at roughly half the branon mass with a continuum of lower energy positrons produced by other gauge boson decay channels.

As for gamma rays, there is also the possibility that instead of coming from decays of hadrons, positrons may be produced by direct branon annihilation into electronn-positron pairs thereby producing a line source of positrons located at an energy equal to the branon mass that cannot be explained by any other standard mechanism. However the $s$-wave annihilation channel is suppressed by the electron mass.

\section*{Antinuclei}
For these particles, the background is the smallest one in comparison to the rest of particles coming from cosmic rays at low energies. For the antiproton case, the background is due to secondary anti-protons produced by cosmic ray collisions with interstellar gas ,where the most common interactions are $p-p$ and $p-He$. It is expected that the background flux of antiprotons should fall dramatically at low energies $T_{\overline{p}}\lesssim \text{GeV}$. Nevertheless branon annihilations are able to produce low-energy antiprotons (branons annihilate in quarks, leptons and gauge bosons who may hadronize into antiprotons whose energy spectrum is determined by fragmentation functions) and therefore observation of low-energy cosmic ray anti-protons could give the best evidence for a branon halo. However, recent experiments such as $\text{BESS}$ collaboration seem to exclude a much higher antiproton flux at those energies than what is predicted through standard cosmic-ray production processes. This fact could still be consistent with a branon halo if $M$ is high enough to produce a non-detectable signal at low energies. The production rate of antiprotons, considering all possible annihilation channels $i$, with $B_{i}$ branching fractions, can be written, following for instance \cite{Antiprotons}, as follows
\begin{eqnarray}
Q_{\overline{p}}(T_{\overline{p}}, r)\,=\,\sum_{f}B_{i}\langle\sigma v\rangle\frac{\text{d}N^{i}_{\overline{p}}}{\text{d}T_{\overline{p}}}\Big(\frac{\rho_{\text{DM}}(r)}{M}\Big)^{2}\, ,
\label{production_rate_antiprotons}
\end{eqnarray}
where $T$ is the $\overline{p}$ kinetic energy and $r$ is the position distance between the production point and the galactic center. Let remind that as branons annihilate in pairs the above rate is proportional to the square of the branon number density  $\rho_{\text{DM}}/M$. 

Taking into account the production rate given at \eqref{production_rate_antiprotons}, and the propagation of the produced antiprotons inside our galaxy, the following expression for the antiproton flux from branon annihilation in the galactic halo may be obtained (see 
\cite{Antiprotons} and references therein):
\begin{eqnarray}
\Phi_{\overline{p}}(r_{0},T)\,=\,C(T_{\overline{p}})\Big(\frac{\rho_{\text{WIMP}}(r_{0})}{M}\Big)^{2}\sum_{f}B_{f}\langle\sigma v\rangle_{f}\Big(\frac{\text{d}N_{\overline{p}}}{\text{d}T_{\overline{p}}}\Big)_{f}\, ,
\end{eqnarray}
where the quantity $C(T_{\overline{p}})$ has the dimension of length divided by solid angle and it is usually defined as
\begin{eqnarray}
C(T_{\overline{p}})\,=\,\frac{1}{4\pi}v_{\overline{p}}\psi_{\overline{p}}^{eff}(\odot,T_{\overline{p}})\, ,
\end{eqnarray}
with the effective energy distribution $\psi_{\overline{p}}^{eff}$  taken at the solar circle.

The evaluation and prospects for the detection of other antinuclei fluxes coming from branon annihilation is similar to the antiproton case. However, it is interesting to note that antideuterons will provide the largest flux for energies below $\text{GeV}$ than for any other antimatter sources. The reason is that whereas antiprotons or positrons production from $Z$ decays is very similar to that from $W$ decays, the creation of low energy antineutrons (and antideuterons consequently) from $ZZ$ is remarkably enhanced, thus giving rise to the mentioned peak in the antideuterons low-energy flux.

\section*{Conclusions}

From very general assumptions in flexible braneworlds, it is possible to study the branon production at different collider experiments with a typical signature given by missing energy and missing $P_T$. These analyses have constrained the model parameters: $f$ and $M$, for different number of branons. Further constraints can also be obtained by computing the
effect of virtual branons on various precision observables
including the muon $g-2$ measurements. Taking all this into 
account, one can estimate the production of cosmic rays coming from 
branon annihilation in the galactic halo, the Sun or the Earth. The final products of these annihilations such as gamma rays, neutrinos, positrons or antimatter in general, may be detected in different experiments providing first evidences of these scenarios.

\begin{acknowledgments}
This work has been partially supported by
DGICYT (Spain) under project numbers FPA 2004-02602 and FPA
2005-02327. The work of AdlCD is also supported by scholarship AP2004-2950 FPU, Ministry of Science, Spain. The work of JARC is supported in part by NSF CAREER grant 
No.~PHY--0239817, NASA Grant No.~NNG05GG44G, the Alfred 
P.~Sloan Foundation and a Gary McCue Postdoctoral Fellowship 
through the Center for Cosmology at UC 
Irvine.
\end{acknowledgments}

\end{document}